\documentclass[oldversion]{aa}
\usepackage{graphicx}
\usepackage{txfonts}
\def\farcs{\hbox{$.\!\!^{\prime\prime}$}}
\def\fdg{\hbox{$.\!\!^\circ$}}

\begin{document}

   \title{The host galaxy of GRB\,011121: Morphology and Spectral Energy Distribution 
     \thanks{Based on observations made with the NASA/ESA Hubble Space 
Telescope under program with proposal ID 9180,
obtained from the data archive at the Space Telescope Science Institute.
STScI is operated by the Association of Universities for 
Research in Astronomy, Inc. under NASA contract NAS 5-26555}\fnmsep
     \thanks{Based on observations made with ESO Telescopes at the La Silla
or Paranal Observatories under program ID 165H.-0464}}

   \author{A. K\"upc\"u Yolda\c{s}
          \inst{1}
          \and
          M. Salvato \inst{1,2}
	  \and
	  J. Greiner \inst{1}
	  \and
	  D. Pierini \inst{1} 
	  \and
	  E. Pian \inst{3} 
	  \and
	  A. Rau \inst{1,2,3}
          }

   \offprints{A. K\"upc\"u Yolda\c{s}, \email{ayoldas@mpe.mpg.de}}

   \institute{Max-Planck-Institut f\"ur extraterrestrische Physik,
              Giessenbachstrasse 1, D-85748 Garching, Germany
	\and Department of Astronomy, California Institute of Technology, 1200 E 		California Blvd, Pasadena, CA, 91125, USA         
	\and INAF - Osservatorio Astronomico di Trieste, via G.B. Tiepolo 
		11, 34131 Trieste, Italy
             }
\date{Received  / Accepted 14 November 2006}

\abstract
{We present a detailed study of the host galaxy of GRB\,011121
(at $z = 0.36$) based on high-resolution imaging in 5 broad-band, optical
and near-infrared filters with HST and VLT/ISAAC. 
The surface
brightness profile of this galaxy is best fitted by a Sersic law with index
$n \sim 2$ -- 2.5 and a rather large effective radius ($\sim$ 7.5 kpc).
Both the morphological analysis and the F450W - F702W colour image suggest
that the host galaxy of GRB\,011121 is either a disk-system with a rather
small bulge, or one hosting a central, dust-enshrouded starburst.  Hence,
we modeled the integrated spectral energy distribution of this galaxy
by combining stellar population and radiative transfer models, assuming
properties representative of nearby starburst or normal star-forming, Sbc-like
galaxies.
A range of plausible fitting solutions indicates that the host galaxy
of GRB\,011121 has a stellar mass of 3.1 -- $\rm 6.9 \times 10^9~M_{\sun}$,
stellar populations with a maximum age ranging from 0.4 to 2 Gyr,
and a metallicity ranging from 1 to 29 per cent of the solar value,
as a function of the time elapsed since star formation started. As for
the opacity, starburst models suggest this galaxy to be nearly as opaque as
local starbursts (with an $\rm A_V = 0.27$ -- 0.76 mag). On the other hand,
normal star-forming Sbc-like models suggest a central opacity larger than that
of local disks by up to a factor of 8, whereas the attenuation along the line
of sight is only $\rm A_V = 0.12$ -- 0.57 mag owing to the galaxy's low
inclination. For this subluminous galaxy (with $\rm L_B/L^{\star}_B = 0.26$),
we determine a model-dependent star formation rate (SFR) of
2.4 -- $\rm 9.4~M_{\sun}~yr^{-1}$, which gives a SFR per unit luminosity of
9.2 -- $\rm 36.1~M_{\sun}~yr^{-1}~(L_B/L^{\star}_B)^{-1}$
and a SFR per unit stellar mass of 0.4 --$\rm 2.9 \times 10^{-9}~yr^{-1}$.
The former specific SFR is high compared to those of most GRB host galaxies,
but consistent with those of most of the hosts at similar low redshift. 
Our results suggest that the host galaxy of GRB\,011121 is a rather large disk-system in a relatively early phase of its star formation history.
   }
   
   \titlerunning{The host galaxy of GRB\,011121}
   \maketitle

%

\section{Introduction}

\subsection{GRB host galaxies}

For nearly all localized Gamma-ray bursts (GRBs) an underlying galaxy
was detected after the decay of the optical/near-infrared (IR) afterglow.
The current sample of long duration GRB (LGRB) host galaxies consists
of $\sim$80 members spanning a large range in magnitudes, i.e. 22 -- 28 mag
in R-band. The observed redshifts of the current sample ranges
from $z = 0.0085$ (Fynbo et al. \cite{fyn00}) to $z = 6.29$
(Berger et al. \cite{ber06})\footnote{see also Jochen Greiner's web page:
http://www.mpe.mpg.de/$\sim$jcg/grb.html}.

The analysis of the observed $\rm R - K$ colour of a subsample of GRB host
galaxies detected until 2002, showed that these are faint blue galaxies
with $\rm R - K = 2.5$ mag in agreement with their nature
of star-forming galaxies (Le Floch et al. \cite{lef03}). The blue colours
of GRB host galaxies are indicators of the link between GRBs and massive-star
formation. Other indicators of the GRB -- massive star connection are
Wolf-Rayet-star signatures (Mirabal et al. \cite{mir03}) and the offsets
between the locations of the GRBs and their host galaxy centers
(Bloom et al. \cite{blo02a}; Fruchter et al. \cite{fr06}).
For four GRBs, the connection between the GRB and the death of a massive star
has been proven unambiguously by the spectroscopic detection of a supernova underlying
the GRB afterglow (Galama et al. \cite{gal98}; Hjorth et al. \cite{hjor03};
Matheson et al. \cite{mat03}; Stanek et al. \cite{sta03}; Malesani et al.
\cite{male04}; Mirabal et al. \cite{mir06}; Pian et al. \cite{pian06}).
Recent studies conclude that the specific star formation rate (SSFR),
i.e the SFR per unit stellar mass, is particularly high for GRB host galaxies,
indicating that they are among the most efficiently star-forming objects
in the universe (Courty et al. \cite{cour04}; Christensen et al.
\cite{chri04b}; Gorosabel et al. \cite{gor05}).

Accurate studies of the morphology, stellar populations, SFRs, and masses 
of GRB host galaxies are obviously ideally conducted at low redshift,
given the better S/N and angular resolution. Photometric and spectroscopic studies of a number of nearby LGRB hosts 
allowed to explore the fundamental characteristics (luminosity, age, 
intrinsic extinction, SFR, metallicity) of those galaxies and has proven 
that detailed host investigations provide important information on the close 
environment of the GRB explosion site (Fynbo et al. \cite{fyn00}; Sollerman et al. \cite{sol05}; Rau et al. \cite{rsg06}).

In general, the faintness of the GRB host galaxies represents a limit
for good S/N spectroscopy. Broad-band spectral energy distributions (SEDs)
are effective substitutes of spectra for determining the galaxy properties.
Analysis of the optical/near-IR SEDs of 11 GRB host galaxies revealed that
the majority are best fitted with starburst galaxy templates (Sokolov et al.
\cite{so01}) using stellar-population models from P\'EGASE (Fioc \& Rocca-Volmerange \cite{f97}) or again with a starburst type galaxy template (Gorosabel et al. \cite{go03a,go03b}; Christensen et al. \cite{chri04a}) of Bruzal \& Charlot (\cite{bc93}) using HyperZ (Bolzonella et al. \cite{bol00}).
This, together with the optical faintness and colours, was recognized as
an indication that long duration GRBs with a detected afterglow
predominantly trace unobscured star-formation in subluminous blue galaxies.

\begin{table*}
\caption{Log of observations and morphological parameters}
\label{tab:obs}
\centering
\begin{tabular}{cccccccc}
\hline\hline
Filter & Date & Tele/Instr & Exposure & Sersic index n & Effective radius & Position angle & Ellipticity$^{\mathrm{1}}$ \\
& & & sec & & kpc & degree & \\
\hline
F450W & 2002-04-21 (day 161) & HST/WFPC2 & 4500 & 2.1$\pm$0.3 & 7.4$\pm$1.4 & 30.7$\pm$2.9 & 
0.52$\pm$0.03  \\
F555W & 2002-05-02 (day 172) & HST/WFPC2 & 4500 & 1.8$\pm$0.1 & 7.2$\pm$0.5 & 31.6$\pm$7.5 & 
0.13$\pm$0.02 \\
F702W & 2002-04-29 (day 169) & HST/WFPC2 & 4500 & 2.7$\pm$0.1 & 9.3$\pm$0.6 & 27.5$\pm$3.0 & 
0.15$\pm$0.01 \\
F814W & 2002-04-29 (day 169) & HST/WFPC2 & 4500 & 2.4$\pm$0.1 & 7.6$\pm$0.5 & 20.6$\pm$4.8 & 
0.13$\pm$0.02 \\
J$_s$ & 2002-02-09 (day 90) & VLT/ISAAC & 1800 & 1.0$\pm$0.5 & 3.9$\pm$2.2 & 19$^{\mathrm{2}}$ & 0.12$^{\mathrm{3}}$ \\
\hline 
\end{tabular}
\begin{list}{}{}
\item[$^{\mathrm{1}}$] Defined as 1 - (semi-minor-axis/semi-major-axis).
\item[$^{\mathrm{2}}$] The best-fit position angle value with an upper limit of 135$\degr$.
\item[$^{\mathrm{3}}$] The best-fit ellipticity value with an upper limit of 0.42.
\end{list} 
\end{table*}

\subsection{GRB 011121}

GRB\,011121 was detected by the Gamma-ray Burst Monitor/Wide-field Camera 
on board {\it BeppoSAX} on 2001 November 21, 18:47:21 UT (Piro \cite{pir01}).
Piro et al. (\cite{piro05}) suggested that there is absorbing gas associated with a star-forming region within a few parsec around the burst in connection with a decreasing column density from N$_H$ = 7$\pm$2$\times$10$^{22}$cm$^{-2}$ to zero during the early phase of the prompt emission.
The optical/near-IR afterglow was discovered independently by several groups
(e.g., Wyrzykowski et al. \cite{wyr01}; Greiner et al. \cite{g01}). Further 
observations revealed excess emission in the light curve associated with a 
supernova (Bloom et al. \cite{blo02b}; Price et al. \cite{pri02};
Garnavich et al. \cite{gar03}; Greiner et al. \cite{g03}).
The spectroscopic redshift of GRB\,011121 is $z$=0.362 from Greiner et al. 
(\cite{g03}) who determined it by fitting the strong host emission lines, i.e. 
H$\alpha$, H$\beta$, [OII], [OIII], underlying the spectrum of the afterglow. 

The host galaxy of GRB 011121 is one of the most extensively and deeply imaged
hosts. High resolution images
are available in optical and near-IR filters covering the rest-frame
wavelength range of $\sim 3200$ -- $\rm 8000~\AA$. This gives us
the unique possibility to study the host galaxy properties through
the parameter space from morphology to stellar mass.

Here we present the morphological and spectral energy distribution analysis
of the host galaxy of GRB\,011121 using archival HST/WFPC2 and VLT/ISAAC data.
In Sections 2, 3 and 4 we present the data reduction, morphological analysis
and the photometry of this galaxy, respectively. In Sect. 5 we analyse
the spectral energy distribution of the host galaxy and derive properties
of the stellar population and the interstellar medium (ISM). In Sect. 6
we calculate the SFR and SSFR and compare the values with other galaxies.
Finally, we summarize our results in Sect. 7.

We adopt $\Omega_\Lambda$ = 0.7, $\Omega_M$ = 0.3
and H$_0$ = 65 km s$^{-1}$ Mpc$^{-1}$ throughout this paper.
The luminosity distance at the redshift of the host ($z = 0.362$)
is D$_L$ = 2080.2 Mpc, and 1 arcsecond corresponds to 5.43 kpc.

\section{Observations and data}

\subsection{Data reduction}

Imaging of the field of GRB\,011121 has been performed at many epochs.
For the present analysis we have chosen the data acquired by
the HST Wide Field Planetary Camera 2 (WFPC2)
and the VLT Infrared Spectrometer And Array Camera (ISAAC), sufficiently late
after the GRB so that the afterglow does not contribute significantly
to the brightness of the host galaxy. The HST data were acquired approximately 5 months after the burst,
using 4 filters: F450W, F555W, F702W and F814W (see Tab.\ref{tab:obs}). These data were obtained as a part of a large program (ID: 9180, PI: Kulkarni) intended to probe the environment of GRBs. 
The total exposure time in each filter is 4500 seconds. An independent
analysis of these data has been published in Bloom et al. (\cite{blo02b}), concentrating on the supernova signature underlying the afterglow lightcurve.

The HST imaging data were pre-processed via ``on the fly'' calibration,
i.e. with the best bias, dark, and flat-field available at the time
of retrieval from the archive. The Wide Field (WF) chips of WFPC2 have a pixel
scale of 0$\farcs$1/pixel. The images for each filter were dithered by
subpixel offsets (resulting in a pixel scale of 0\farcs05/pixel) using
the IRAF/Dither2 package to remove cosmic rays and produce a better-sampled
final image. For all HST observations, the host position falls near the serial
readout register of WF chip 3 which minimizes the correction for charge
transfer efficiency (CTE) to around 5 per cent in count rate, therefore
we ignore the CTE correction for the photometry. 

The VLT/ISAAC data were obtained in the $\rm J_s$-band on February 9, 2002
with an exposure of 1800 seconds (see Tab.\ref{tab:obs}), and reported earlier
in Greiner et al. (\cite{g03}). These data were also obtained as a part of a large program (ID: 165H.-0464, PI: van den Heuvel) intended to understand the physics of GRBs and the nature of their hosts. The $\rm J_s$-band images were reduced using the ESO Eclipse package (Devillard \cite{dev05}).

\begin{figure}
\begin{center}
\includegraphics[width=8cm,angle=0]{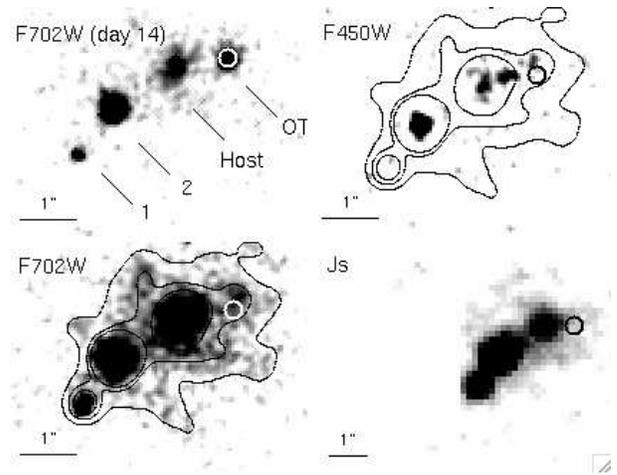}
\caption{{\it Top left}: F702W image taken $\sim$14\,days after the GRB. Two 
foreground stars, the positions of the host, and of the optical afterglow (circle in all panels) are indicated. {\it Top right} and {\it Bottom left:} F450W 
and F702W images taken $\sim$5\,months post-burst. The contours show the light 
distribution in the F702W filter. {\it Bottom right}: J$_s$ image taken 3\,months after the burst. All images are tophat smoothed. North is up and East is to the left.}
\label{fig:hostim}
\end{center}
\end{figure}

Zero-point magnitudes for the HST filters were taken
from Dolphin (2000)\footnote{see also
http://purcell.as.arizona.edu/wfpc2$\_$calib/}.
For the VLT images, two local photometric standard stars given
by Greiner et al. (\cite{g03}) were used to obtain the photometric calibration.
Both for the HST and the VLT data, the background values of the images
were calculated using IRAF/imexamine in the corresponding filters.
The 1$\sigma$ surface brightness limits are calculated using the formula
given by Temporin (\cite{tem01}):
\begin{equation}
\mu_{lim} =  -2.5 \times log[\sigma/(t \times s^2)] + \mu_0
\end{equation}
where $\sigma$ is the standard deviation from the mean of the background, 
$\mu_0$ is the zero-point, $t$ is the exposure time in seconds
and $s$ is the pixel scale. 

\subsection{Astrometry}

Images obtained at different epochs and different filters were registered relatively to an early F702W image where the OT is clearly visible (top left image of Figure\,\ref{fig:hostim}), using standard MIDAS routines. We used at least three isolated stars to find the relative shift and rotation of two images. The centers of the stars were computed assuming a point source. We did not re-scale the images since the HST images have
the same scale. 
The estimated accuracy of our relative astrometry is 10 mas given by the rms error of the mapping using MIDAS routines. We note that the uncertainties due to optical distortion for the HST images are rather small and are largely removed by the dithering process (Fruchter \& Hook \cite{fruh02}).  The relative position of the OT in the $J_{s}-band$, as shown in the bottom right image of Figure\,\ref{fig:hostim}, is similarly estimated using an early VLT/ISAAC $J_{s}-band$ image from Nov 24, 2001 (see Greiner et al. \cite{g03}), with an rms of 30 mas.

\subsection{Extinction}

As for the necessary correction for Galactic extinction, the study
of Schlegel et al. (\cite{schl98}), based on COBE and IRAS extinction maps,
gives a value of Galactic reddening along the line of sight of GRB\,011121
equal to $\rm E(B-V) = 0.49$ mag. However, different authors have argued that
extinction estimates based on far-IR measurements overpredict the true value
by about 30\% (Dutra et al. \cite{dut03}; Cambr\'{e}sy et al. \cite{cam05}).
In particular, Dutra et al. (\cite{dut03}) recommend to scale the value
of $\rm E(B-V)$ given by Schlegel et al. (\cite{schl98}) by a factor of 0.75
for lines of sight corresponding to regions with $\rm |b| < 25^o$
and $\rm E(B-V) > 0.25$ mag. This holds for the line of sight of GRB\,011121,
hence we assume $\rm E(B-V) = 0.37$ mag as the correct value of Galactic
reddening. This value corresponds to a V-band extinction $\rm A_V = 1.15$ mag
for the standard Galactic extinction curve of Cardelli et al. (\cite{car89}), where
$\rm R = A_V / E(B-V) = 3.1$. We correct the observed photometry of
the host-galaxy of GRB\,011121 for Galactic extinction according to this law.

Using the broad-band spectral energy distribution of the optical transient (OT) of GRB\,011121, Garnavich et al. (\cite{gar03}) estimated $\rm E(B-V) = 0.43 \pm 0.07$ mag, and Price et al. (\cite{pri02}) estimated $\rm A_V = 1.16 \pm 0.25$ mag for the {\it total} (i.e. Galactic plus internal) reddening.
These two analyses offer
consistent results as for the {\it total} extinction and reddening,
within the uncertainties. However, note that these authors implicitly assumed
that the solution of radiative transfer for the light through 
the host-galaxy of GRB\,011121 is the same as for the light from a star
in the Galaxy.

Our assumed values of Galactic reddening and extinction are consistent with
the previous {\it total} values, within 1 $\sigma$. However, we do not
conclude that the extinction produced by dust in the host-galaxy
of GRB\,011121 is negligible.
In fact, the optical spectra of two slightly different regions (due to different slit widths) containing
the OT of GRB\,011121, taken by Greiner et al. (\cite{g03}) 4 and 21 days
after the GRB event, give values of the Balmer-line flux ratio
$\rm H_{\alpha} / H_{\beta}$ equal to $4.8^{+1.6}_{-1.1}$ and $6.4^{+3.5}_{-1.9}$,
respectively, after correcting the line fluxes for foreground extinction.
Both Balmer-line flux ratios derived from Greiner et al. (\cite{g03})
are higher (by $>$ 2 $\sigma$) than the value of 2.86 predicted for
the standard case B recombination\footnote{Although the blast wave of the GRB
may cause shock-ionization, Perna et al. \cite{prl} showed that it is expected
to influence the ionization state of the gas on timescales of hundreds
to thousands of years after the burst. Therefore we take the case B
recombination as representative of the dust-free case, and assume that
the photo-ionization effect of GRB prompt and afterglow emission
on the circumburst environment is negligible (see K\"upc\"u Yolda\c{s} et al.
\cite{aky}). (e.g. Osterbrock \cite{os89}) and implies an A$_V$ of 1.6$^{+0.9}_{-0.8}$ and 2.5$^{+1.4}_{-1.9}$, respectively, derived using the extinction curve of Cardelli et al. (\cite{car89}). Higher than predicted Balmer-line
flux ratios are due to dust present in the small-/large-scale environment
of H\,{\small II} regions (Cox \& Mathews \cite{cm69}; Mathis \cite{m70}).
Hence the presence of a non-negligible amount of dust extinction
in the host-galaxy of GRB\,011121 is a feasible working hypothesis.}

\section{Morphology of the host galaxy}

\begin{figure}
\includegraphics[width=8cm,angle=0]{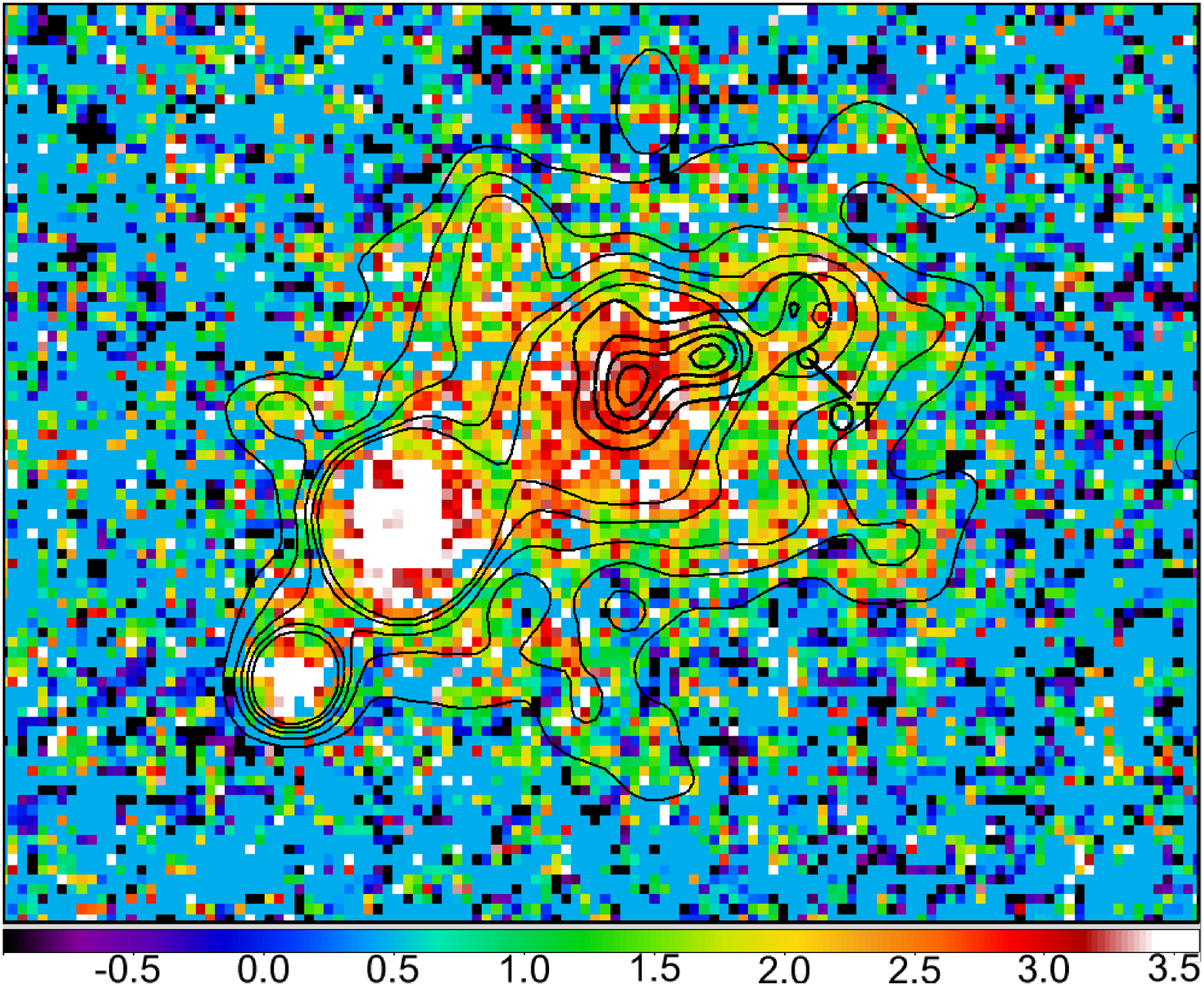}
\caption{F450W -- F702W color image of the field of GRB 011121. The position of the OT is indicated with an arrow. The thin-line contour is 
the the contour of the galaxy in the F702W filter and the thick-line is 
the contour in the F450W filter, overplotted on the color image.}
\label{fig:bminr}
\end{figure}

The high-resolution data in 5 broad-band filters allow
a colour-resolved morphological analysis. Figure \ref{fig:hostim} shows
images of the host galaxy of GRB\,011121 in various filters.
This galaxy exhibits a different structure in the F450W band compared to
the redder band data (see Fig.~\ref{fig:hostim} top right and bottom left
images). In the F702W image we see a nearly face-on extended structure.
On the other hand, the F450W image -- despite the lower sensitivity -- reveals three emission regions
, most probably indicating the sites of enhanced star 
formation in the galaxy, considering that the size of a star forming region  ($\sim$ few pc) is much smaller than the sizes of these blue emission regions ($\sim$1-2 kpc). The difference of morphology in different filters 
is reflected in the F450W -- F702W color image of the galaxy
(see Fig.~\ref{fig:bminr}). The center of the galaxy is red
with $\rm F450W - F702W = 3.0 \pm 0.1$ mag, the background value being
$\rm F450W - F702W = 0.2 \pm 0.2$ mag. The three emission regions seen
in the F450W filter exhibit $\rm F450W - F702W$ equal to 2.6$\pm$0.1 mag,
1.5$\pm$0.1 mag and 0.95$\pm$0.15 mag, respectively. 

\begin{figure}
\begin{center}
\includegraphics[width=9cm,angle=0]{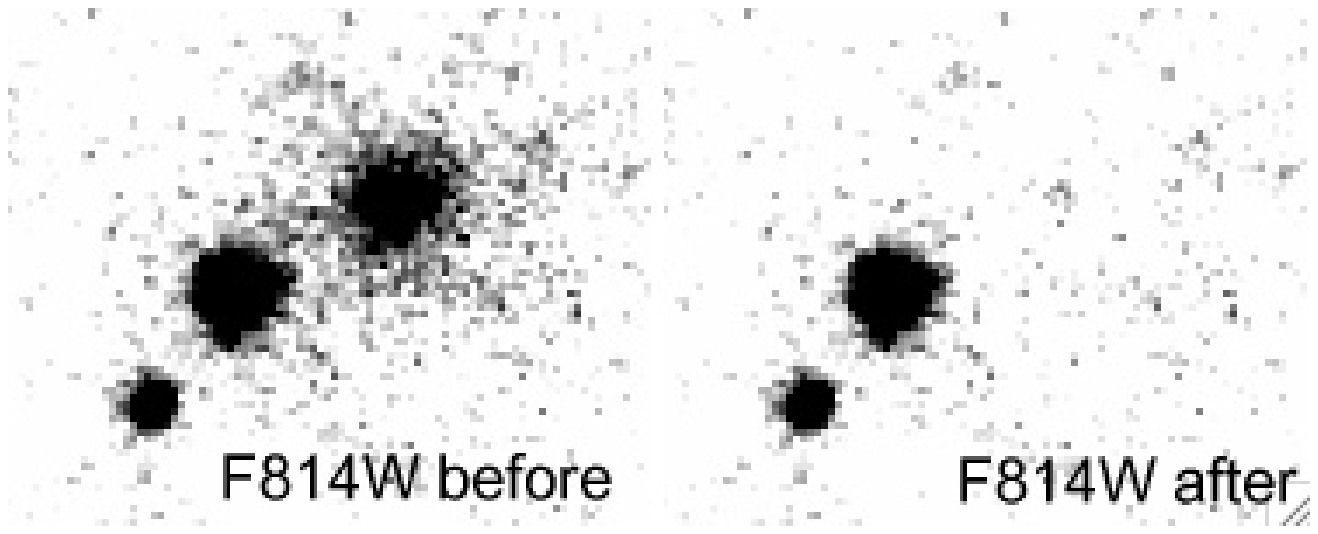}
\includegraphics[width=4.5cm,angle=-90]{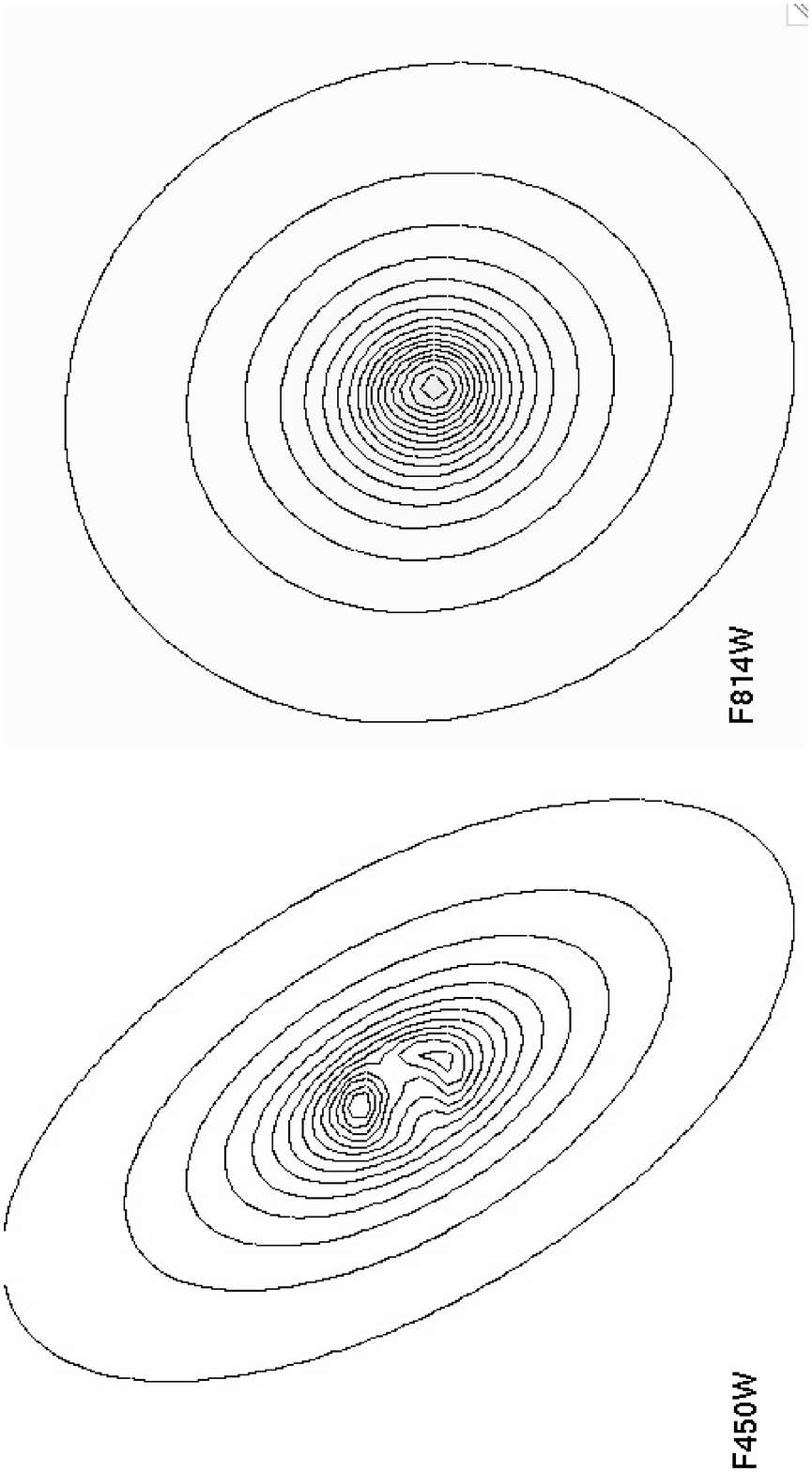}
\caption{{\it Top Panel Left}: The image of the field in the F814W filter in April 2002. {\it Top Panel Right}: The residuals after subtracting the best-fit Galfit galaxy model from the original image. {\it Bottom Panel} The contours of the best-fit model of the Galfit analysis for the F450W data (on the {\it left}) and for the F814W data (on the {\it right}).}
\label{fig:galfres}
\end{center}
\end{figure}

The morphological analysis of the host galaxy of GRB\,011121 was performed
using Galfit (Peng et al. \cite{peng02}). Galfit is a 2D galaxy
and point-source fitting algorithm which can fit an image with multiple
analytical models simultaneously. For the galaxy under investigation,
an initial model assuming a classical de Vaucouleurs bulge$+$exponential disk
profile did not provide a good representation. Therefore, we made use
of a Sersic profile (Sersic \cite{sers}) where all the related parameters
(i.e. effective radius, Sersic index, position angle) were left free.
The top panel of Figure \ref{fig:galfres} shows the image of the field of the host galaxy 
in the F814W band, and the residual image after the subtraction of the 
galaxy model. The results of the best fits obtained with Galfit for 
each filter are listed in Table \ref{tab:obs}. 

The best-fit values 
for ellipticity and position angle are in agreement with each other
for all filters, except the ellipticity for the F450W filter (see the bottom panel of Fig.\,\ref{fig:galfres}). There is a similar agreement for the effective radius and the Sersic index parameters.
We note that the values for the F450W fit should be evaluated carefully, considering that the galaxy image has
a relatively lower signal-to-noise ratio due to the sensitivity
of the detector and therefore probably probes only the high surface brightness regions. Nevertheless, the values except the ellipticity are still in agreement for all images, indicating that we actually trace the profile of the galaxy in a decent way. 

\begin{figure}
\includegraphics[width=8cm,angle=0]{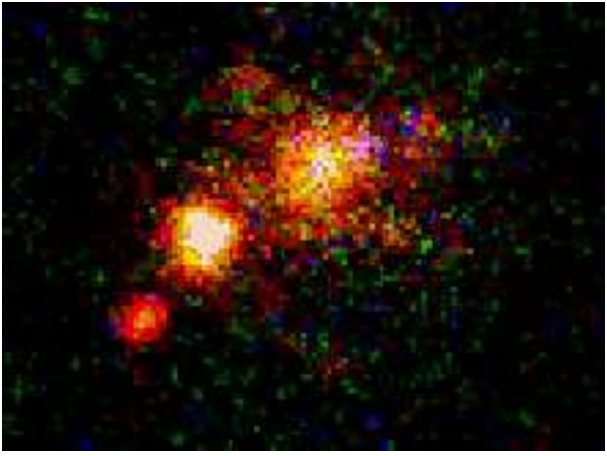}
\caption{A white image of the field of GRB 011121 constructed using the images in the F450W ({\it blue}), F555W ({\it green}), F702W and F814W ({\it red}) filters.}
\label{fig:white}
\end{figure}

Galaxies at cosmological redshifts are commonly classified according to their Sersic index as disk systems ($n < 2$) and bulge-dominated systems ($n > 2$, see Ravindranath et al. \cite{rav04}). However, we note that a central, dust-enshrouded starburst can produce a Sersic profile with index
of about 2 and a redder $\rm F450W - F702W$ colour in the inner region
of a disk system as seen for the host of GRB 011121 (see Fig.\ref{fig:bminr}).
The detection of a bulge can be hindered by the fact that the galaxy
is observed nearly face-on, the best-fit ellipticity value being 0.13
(0.50 for F450W). Although the Sersic index of our reddest band data ($\rm J_s$-band) is consistent with values typical of a disk-dominated galaxy, this is still consistent with an extended disk structure dominating a small, unresolved
bulge, since the spatial resolution of the $\rm J_s$-band image is almost
three times worse than that of the HST images. We also inspected
the $\rm F555W - J_s$ radial colour profile and found that it is constant
within the errors, indicating that there is no significant difference
in the radial profile of the galaxy in different filters except for F450W.
Therefore, the host galaxy of GRB\,011121 can be
either a disk system with a small bulge as also indicated by the enhanced traces of spiral arms in Figure\,\ref{fig:white}, i.e. an Sbc-like galaxy, or a disk system experiencing dust-enshrouded starburst activity
in its central regions.

Similar results on the morphology of the host galaxy of GRB\,011121
were obtained by two other groups using different methods. Wainwright et al.
(\cite{wain05}) performed a morphological analysis using Galfit on the same HST 
data as used here plus the F850L filter data; they concluded that the galaxy
is a disk system. Our results are generally in agreement with those
of Wainwright et al. (\cite{wain05}), except for the F450W filter,
for which there is a $\sim$4$\sigma$ difference in the effective radius. Note that we cannot quantify the difference since Wainwright et al. did not quote any errors for their results.
On the other hand,
also Conselice et al. (\cite{con05}) performed a morphological analysis based
on the concentration and asymmetry parameters using the F702W filter data
taken $\sim$3 months after the GRB. They concluded that the host is probably
a late-type spiral consistent with our results.

The OT of GRB\,011121 was clearly distinguishable in earlier images taken with HST/WFPC2 since it is located in the outskirts of its host galaxy (top left image of Fig.\ref{fig:hostim}). None of the emission regions seen in the F450W band data coincides with the OT position (see the top right image Fig.\ref{fig:hostim}).

In addition, we investigated the nature of the two objects in the vicinity
of the host galaxy. The radial surface brightness profile of these objects
is described by the point spread function in the HST images, as estimated
from the stars in the field. Furthermore, there was no X-ray emission
associated with these objects in the X-ray imaging of the afterglow. Hence
we conclude that the objects marked as number 1 and 2
in Figure \ref{fig:hostim} (top left) are most probably foreground stars. We conducted the photometry of these objects including also the H and K data from Nov 24, 2001 (ID: 165H.-0464, PI: van den Heuvel) acquired by VLT/ISAAC, in order to estimate the spectral type assuming that they are stars. The colors of object 2 are $V-R$=1.16$\pm$0.10 mag, $J-H$=0.62$\pm$0.05 mag and $H-K$= 0.14$\pm$0.03 mag. These colors indicate that object 2 is a main-sequence star of spectral type of M2 (Tokunaga \cite{tok00}). The colors of object 1 are much redder with $V-R$=2.85$\pm$0.10 mag, $J-H$=0.17$\pm$0.10 mag and $H-K$=0.61$\pm$0.12 mag. These colors fit marginally with that of a late M-type or an early L-type star (Tokunaga \cite{tok00}; Leggett et al. \cite{leg03}). However, we do not exclude the possibility that object 1 may be an unresolved high-redshift galaxy. 

\section{Photometry}

Photometry was extracted using the IRAF/Ellipse task which performs
aperture photometry inside elliptical isophotes. To determine the size
of an aperture which covers the galaxy and minimizes the contamination
by the background noise, the 1$\sigma$ surface brightness limit
and the metric radius were calculated for each image. The metric radius
is defined as the radius where the Petrosian index $\eta$ = 0.2,
the Petrosian index being the ratio of the average surface brightness
within a radius $r$ to the surface brightness at $r$ (Petrosian \cite{pet76};
Djorgovski \& Spinrad \cite{ds81}). Both values correspond to a semi-major
axis length of 2.1 -- 2.4 arcsec for all images except for the F450W filter
image for which the surface brightness limit is reached at $\sim$1$\arcsec$.
In order to conduct a consistent analysis, we performed aperture photometry
on each image with the same semi-major axis length of 2.25 arcseconds.
Table \ref{tab:phot} shows the resulting magnitudes and errors. The errors
in magnitudes were calculated assuming Poisson noise and include the readout
noise and zero-point errors. The background fluctuation values were obtained
by calculating the standard deviation from the mean background values
measured for several different areas near the galaxy. Then a correction due
to dithering was applied to the background noise of the HST images, assuming
that the dither pattern is uniform (see Fruchter \& Hook \cite{fruh02}). 

Magnitudes were computed using i) the best-fit
ellipticity and position angle for each filter obtained by Galfit,
and ii) fixing the ellipticity and position angle to 0.13 and 27\fdg5,
respectively for all filters. The results were the same for both cases.
Ellipse also provides the magnitudes inside a circular area
having the same radius of the semi-major axis of the elliptical isophote. 
We compared the magnitudes determined within the circular and elliptical areas
and found that the difference is $<$0.02 mag. This indicates the reliability 
of the 2$\farcs$25 extent, the position angle and the ellipticity of the 
galaxy.

\begin{table*}
\caption{Results of the Photometry}
\label{tab:phot}
\centering
\begin{tabular}{cccc}
\hline\hline
Filter & Brightness$^{\mathrm{1}}$ & Foreground extinction & Absolute 
magnitude$^{\mathrm{2,3}}$ \\
& mag & mag & mag \\
\hline
F450W & 23.44$\pm$0.04 & 1.43 & -19.5 \\
F555W & 22.64$\pm$0.02 & 1.14 & -20.3 \\
F702W & 21.63$\pm$0.01 & 0.86 & -20.6 \\
F814W & 21.18$\pm$0.02 & 0.67 & -21.1 \\
J$_s$ & 19.87$\pm$0.06 & 0.32 & -22.1 \\
\hline
\end{tabular}
\begin{list}{}{}   
\item[$^{\mathrm{1}}$] Magnitudes are not corrected for Galactic extinction.
\item[$^{\mathrm{2}}$] The absolute magnitudes are corrected
for Galactic extinction.
\item[$^{\mathrm{3}}$] The absolute magnitudes are given for the filters 
B, V, R, I, J in respective order.
\end{list}
\end{table*}

The value of M$_B^*$ (uncorrected for dust attenuation)
for redshifts between 0 and 0.5 is given by Dahlen et al.
(\cite{dah05}) as -21.06$^{+0.10}_{-0.06}$ for $h = 0.65$.
It is derived by fitting a Schechter luminosity function and using all types
of galaxies, i.e. early type, late type and starbursts. From this value,
we determine a luminosity ratio of L$_B$/L$^{*}_{B}$ = 0.26
for the host galaxy of GRB\,011121, which indicates that this galaxy
is subluminous.

\begin{figure}
\includegraphics[width=6cm,angle=-90]{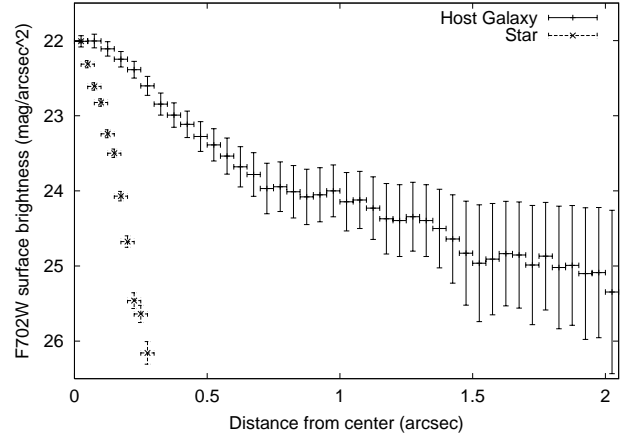}
\caption{The surface brightness profile of the host galaxy and a star in 
the F702W filter. The surface brightness values of the star are 
normalized to those of the galaxy for plotting purposes.}
\label{fig:rprof}
\end{figure}

\section{Analysis of the Spectral Energy Distribution}

Hereafter we analyze the SED of the host galaxy of GRB\,011121
to deduce galaxy properties like characteristic age and metallicity
of the stellar populations and the SFR. We apply both the publicly available HyperZ code (Bolzonella et al. \cite{bol00}) as well as our own modelling, to explore the galaxy properties. 

\subsection{Analysis using HyperZ}

Following the seminal work on GRB host galaxies by Christensen et al.
(\cite{chri04a,chri04b}), we make use of 
{\em HyperZ} (Bolzonella et al. \cite{bol00}).
In particular, this code considers a large grid of models based on 8 different
synthetic star-formation histories (Bruzual \& Charlot \cite{bc93}),
roughly matching the observed properties of local field galaxies (starburst,
elliptical, spiral, and irregular ones). For all models, metallicity is fixed to the solar value ($Z=0.02$).
The empirical formula of Calzetti et al.
(\cite{cal00}) for nearby starbursts is used to describe attenuation by dust
in galaxies, independent of the star-formation history and morphology. Finally,
a Miller \& Scalo (\cite{ms79}) initial mass function with an upper mass limit
for star formation of 125 $\rm M_{\sun}$ is used.

As a result of the fitting of the broad-band photometry of the host galaxy
of GRB\,011121 with {\em HyperZ} models, we find that old ages
(i.e. $\ge$ 1 -- 2 Gyr) are not favoured (best-fit values of 45 Myr
for starbursts and up to 720 Myr for spirals and irregulars), while
the amount of internal extinction is non-negligible
($\rm A_V$ = 0.80 -- 1.0 mag, rest frame) for all models producing
equally valid fits with $\chi_{\nu}^2 <  0.26$. For a so-called Calzetti law,
$\rm A_V$ = 0.80 -- 1.0 mag corresponds to $\rm E(B-V) = 0.20$ -- 0.25 mag.
We note that this value of reddening by internal dust refers to the whole
galaxy and, thus, is not directly comparable in a quantitative way
to the values estimated from spectroscopy of the OT region, once
the contribution of Galactic reddening is removed.

These results hold independent of the synthetic star-formation history
of the model, which mirrors the fact that the 4000 $\AA$-break is not
very prominent in stellar populations younger than $\sim$1 Gyr and, thus,
does not offer a robust constraint to discriminate different evolutionary
patterns. 
 Finally, we note that an even broader range
of possible values for age and extinction exists if we consider fits
with $\chi_{\nu}^2 < 1$. This increase in degeneracy of the solutions is not
a shortcoming of {\em HyperZ} because it was designed to find photometric
redshifts and provides only a rough estimate of the SED type (see Bolzonella
et al. \cite{bol00}), independent of morphology.

\subsection{Broad-band SED fitting}

In order to exploit the information on morphology available
for the host galaxy of GRB\,011121 and better link the mode of star-formation
and the properties of dust attenuation, we build our own set of physically
motivated models. We combine different, composite stellar population models
and models of radiative transfer of the stellar and scattered radiation
through different dusty media. We use a tailored grid of parameters in order
to probe the very wide parameter space available for models in an efficient
way. A large suite of synthetic SEDs is built as a function of total
(gas$+$stars) mass, age (i.e. the time elapsed since the onset of star
formation) and a characteristic opacity of the model, as described in
the following subsections. These three free model parameters are determined
from the comparison of synthetic broad-band apparent magnitudes
(observed frame) and the apparent magnitudes determined for the host galaxy of
GRB\,011121 (see Sect. 4) through the standard least-square fitting technique.

\subsubsection{Stellar population models}

We model the intrinsic (i.e. not attenuated by internal dust) SED
of the host galaxy of GRB\,011121 as a composite stellar population.
We make use of the stellar population evolutionary synthesis code
P\'EGASE (Fioc \& Rocca-Volmerange \cite{f97}) (version 2.0) in order
to compute both the stellar continuum emission and the nebular emission.
Gas is assumed to be transformed into stars of increasing metallicity
as the time elapsed since the onset of star formation increases,
the initial metallicity of the ISM being equal to zero. The stellar initial
mass function (IMF) is Salpeter (\cite{s55}), with lower and upper masses
equal to 0.1 and 120 $\rm M_{\sun}$, respectively. Adopting a different IMF
affects mostly the determination of the stellar mass; for instance,
a Chabrier (\cite{cha03}) IMF produces stellar masses lower by about
30 per cent than a Salpeter (\cite{s55}) one.

The mass-normalized SFR of the models is assumed either to be constant
({\it starburst} models) or to decline exponentially as a function of time
({\it normal star-forming galaxy} models). For models of a normal star-forming
galaxy, we adopt e-folding times equal to 1 and 5 Gyr to describe
the star-formation histories of the bulge and disk components, respectively,
the bulge-to-total mass ratio being set equal to 0.05, 0.1, 0.15 or 0.2.
For starburst models, a range of 18 ages between 0.1 and 9 Gyr
is considered\footnote{Models older than 9 Gyr do not offer a physical
representation of a galaxy at $z = 0.362$ as the host of GRB\,011121.},
the time step being fine (i.e 0.1 Gyr) up to an age of 1 Gyr and coarse
(i.e. 1 Gyr) since then. On the other hand, for normal star-forming galaxy models,
a range of 28 ages between 0.5 and 7 Gyr is considered. For these models, a fine time
step is adopted for ages between 1 and 3 Gyr in order to better follow
the different evolution of the stellar populations of the bulge and disk
components. Finally, we assume that the total mass of the system ranges from
$10^9$ to $\rm 2 \times 10^{11}~M_{\sun}$, 200 steps in mass being considered.

\subsubsection{Dust attenuation models}

As a statistical description of dust attenuation in starbursts,
we make use of the Monte Carlo calculations of radiative transfer
of the stellar and scattered radiation by Witt \& Gordon (\cite{wit00})
for the SHELL geometry. 
In this case, stars are surrounded by a shell where a two-phase clumpy medium
hosts dust grains with an extinction curve like that of the Small Magellanic
Cloud (SMC), as given by Gordon et al. (\cite{gcw97}). We note that these
models describe dust attenuation in nearby starburst galaxies
(Gordon et al. \cite{gcw97}) as well as in Lyman Break Galaxies at $2 < z < 4$
(Vijh et al. \cite{vij03}). We consider 14 values of the opacity $\tau_V$
(0.25 -- 9), where $\tau_V$ is the radial extinction optical depth
from the center to the edge of the dust environment in the V-band, assuming
a constant density, homogeneous distribution.

On the other hand, for the normal star-forming galaxy models
we assume that dust attenuation is described by
the Monte Carlo calculations of radiative transfer
of the stellar and scattered radiation for an axially symmetric disk geometry
illustrated in Pierini et al. (\cite{dp04b}) and based on the DIRTY code
(Gordon et al. \cite{gor01}). These models have been applied successfully
to interpret multiwavelength photometry of edge-on late-type galaxies
in the local Universe (Kuchinski et al. \cite{kuc98}). 
The physical properties of the dust grains are assumed to be the same
as those in the diffuse ISM of the Milky Way (from Witt \& Gordon
\cite{wit00}). Furthermore, this time we use as a parameter
the central opacity $\tau_V^{\rm c, 0}$, that refers to the face-on extinction
optical-depth through the centre of the dusty disk in the V-band. In these
disk models, the central opacity is equal to 0.5, 1, 2, 4, 8, and 16.

From the observed ellipticity of the host galaxy of GRB\,011121
(see Tab. \ref{tab:obs}), we determine an inclination of about 18 degrees,
for an intrinsic axial ratio of 0.2. Hence we adopt disk galaxy models
with only this inclination since inclination effects on the total luminosity
are small for inclinations much less than 70 degrees in a disk-dominated
system (e.g. Pierini et al. \cite{dp03}) like the host galaxy of GRB\,011121.
In fact, the Sersic index fitted to different light profiles
of the host galaxy of GRB\,011121 (see Tab. \ref{tab:obs})
is consistent with the presence of a small bulge like in Sbc galaxies. Greiner et al. (\cite{g03}) estimated
the bulge-to-disk (B/D) $J_s$-band luminosity ratio to be about 0.28
using a de Vaucouleurs$+$exponential model to reproduce the $J_s$-band 
surface brightness profile of the host galaxy of GRB\,011121.
Hence, we use a bulge-to-disk $J_s$-band luminosity ratio between 0.23 and 
0.33 as a further constraint for our bulge$+$disk models allowing for
mismatches between the fitting model of Greiner et al. (\cite{g03})
and the structure of the system described in Pierini et al. (\cite{dp04b}).

Finally, for all models we assume that the gas emission at a given wavelength
is attenuated by the same amount as the stellar emission at that wavelength,
independent of whether the gas emission is in a line or in the continuum
(see Pierini et al. \cite{dp04a} for a discussion).

\subsubsection{Results}

For a suite of 50,400 starburst models plus 124,800 normal star-forming models,
synthetic SEDs and magnitudes are computed and evaluated against the observed
broad-band SED of the host galaxy of GRB\,011121 (see Sect. 4). Reassuringly,
each suite of models brackets the best-fit solution although the parameter space
is not spanned in a uniform way. Hereafter we illustrate the basic aspects
of those fit solutions that are called ``plausible'', being characterized
by $\chi_{\nu}^2 < 6.91$, that corresponds to a probability of 0.001
for two degrees of freedom (given by 5 photometric points minus 3 model
parameters).

\begin{figure} 
\includegraphics[width=9cm,angle=0]{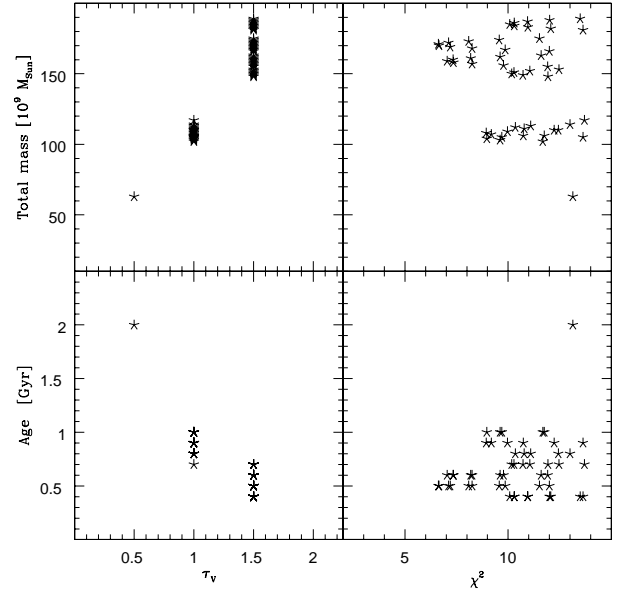}
\caption{SED fit solutions with $\chi_{\nu}^2 < 6.91$, using starburst models.
{\it Left}: Total (gas$+$stars) mass and age versus $\tau_V$, {\it Right}:
Total mass and age versus $\chi_{\nu}^2$.}
\label{fig:fSB}
\end{figure}

As Fig. \ref{fig:fSB} shows, plausible solutions for the starburst case
imply ages between 0.4 and 2 Gyr and, accordingly, an opacity decreasing
from 1.5 to 0.5. This domain is narrower than the explored parameter space,
nevertheless it still expresses the well-known age--opacity degeneracy
for starbursts (Takagi et al. \cite{tak99}). At the same time, the bolometric
luminosity-weighted metallicity in stars increases from $3 \times 10^{-4}$
to $1.6 \times 10^{-3}$, while the total mass of the system drops from 18.5
to $\rm 6.3 \times 10^{10}~M_{\sun}$. The latter range corresponds to a range
of 3.1 -- $\rm 4.8 \times 10^{9}~M_{\sun}$ in stellar mass.
In particular, the best-fit model for the starburst case has an age
of 0.5 Gyr, a bolometric luminosity-weighted metallicity in stars equal
to $3.7 \times 10^{-4}$, a stellar mass of $\rm 3.6 \times 10^{9}~M_{\sun}$
and an opacity equal to 1.5\footnote{The two-phase, clumpy SHELL model
of Witt \& Gordon (\cite{wit00}) with SMC-type dust and $\tau_V = 1.5$
produces an attenuation curve that best matches the so-called Calzetti law
for nearby starbursts (see Calzetti et al. \cite{cal00} and references
therein).}. We note that $\tau_V = 1.5$ corresponds to an attenuation of
the total flux at V-band (rest frame) $\rm A_V = 0.76~mag$
and a reddening $\rm E(B-V) = 0.20~mag$ on the scale of the system.

\begin{figure}
\includegraphics[width=9cm,angle=0]{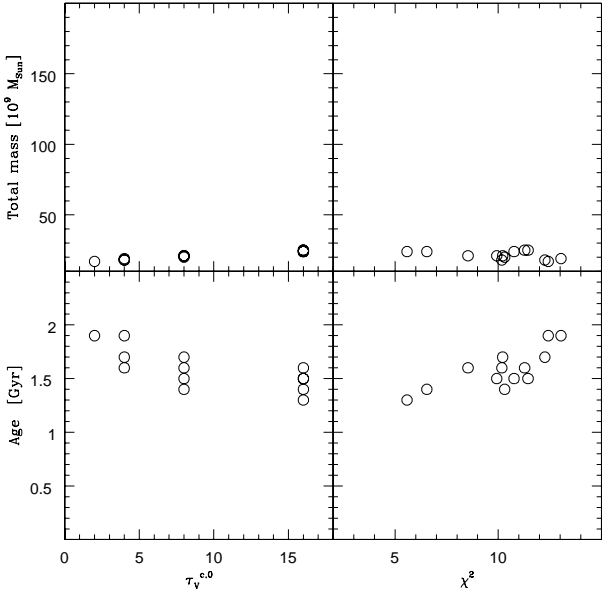}
\caption{Same as Fig.\ref{fig:fSB} for normal star-forming galaxy models.}
\label{fig:fSbc}
\end{figure}

On the other hand, plausible solutions for the normal star-forming case
have a bulge-to-total mass ratio equal to 0.15. They imply ages between 1.3
and 1.9 Gyr and, accordingly, a central opacity of the disk decreasing
from 16 to 2 (see Fig. \ref{fig:fSbc}). At the same time, the bolometric
luminosity-weighted metallicity in stars of the disk increases
from $3.9 \times 10^{-3}$ to $5.8 \times 10^{-3}$. The total mass of the system
drops from 2.5 to $\rm 1.7 \times 10^{10}~M_{\sun}$ from the youngest
and most opaque systems to the oldest and least opaque ones. The range
in stellar mass 
spanned by these plausible solutions is 4.9 -- $\rm 6.9 \times 10^{9}~M_{\sun}$.
In particular, the best-fit model for the normal star-forming case has an age
of 1.3 Gyr, a bolometric luminosity-weighted metallicity in stars of the disk
equal to $3.9 \times 10^{-3}$, a stellar mass of
$\rm 5.7 \times 10^{9}~M_{\sun}$ and a central opacity of the disk equal
to 16. We note that $\tau_V^{\rm c, 0} = 16$ corresponds to an attenuation
(along the line of sight) of the total rest-frame V-band flux
$\rm A_V = 0.57~mag$ for an inclination of 18 degrees. In terms of reddening
of the stellar component of the only disk, the best-fit Sbc-like model implies
$\rm E(B-V) = 0.08~mag$ on the disk scale. Even smaller values of reddening
will apply to a peripheral region of the disk, where the OT of GRB\,011121
was actually located. Hence plausible solutions for a normal star-forming
bulge$+$disk system comfortably meet the constraints on a low amount
of reddening in the OT region of GRB\,011121.

\begin{figure}
\includegraphics[width=9cm,angle=0]{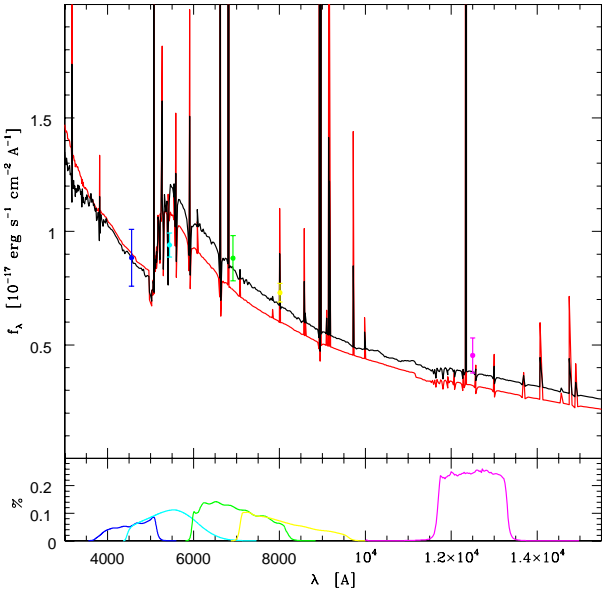}
\caption{The best-fit normal star-forming galaxy model (in black), and the best-fit starburst model (in red). The points are the fluxes of the host galaxy derived from the observed magnitudes corrected for the foreground extinction. The filter curves are shown in the lower panel, for the corresponding filters.}
\label{fig:bestf}
\end{figure}

Figure \ref{fig:bestf} shows how the best-fit models
for a starburst system and a normal star-forming bulge$+$disk system reproduce
the observed photometry of the host galaxy of GRB\,011121.
The comparison with the data reveals that both best-fit models underpredict
the observed $\rm J_s$-band magnitude by about 0.1 mag, i.e. almost 2 $\sigma$.
This is the main reason for their rather high values of $\chi_{\nu}^2$.
A posteriori, we interpret this discrepancy as due to the fact that P\'EGASE
(version 2.0) does not include the contribution to the total emission from
the thermally pulsating asymptotic giant branch (TP-AGB) phase of stellar
evolution (see Maraston \cite{mar05}). TP-AGB stars are cool giants exhibiting
very red optical/NIR colours (e.g. Persson et al. \cite{per83}). They are
expected to play a significant role in the rest-frame visual-to-near-IR
emission of galaxies containing 1-Gyr-old stellar populations
(Maraston \cite{mar98,mar05}). Now the best-fit models contain
stellar populations that are up to 0.5 or 1.3 Gyr old (starburst
or Sbc-like model, respectively), hence it is plausible that
they can slightly underpredict the flux in the observed $\rm J_s$-band
magnitude\footnote{We note that the models of Bruzual \& Charlot (\cite{bc93})
included in {\em HyperZ} (Bolzonella et al. \cite{bol00}) do not include
the contribution to the total emission from the TP-AGB stars
(see Maraston \cite{mar05}) as well. However, they have stellar populations
with only solar metallicity, which are redder than those
with lower metallicity.}.

We tested that the previous results are not biased by the absence
of the contribution to the total emission from the TP-AGB stars
in P\'EGASE (version 2.0). We performed new fits where the range
in the $\rm J_s$-band B/D allowed by the estimate of Greiner et. al.
(\cite{g03}) and/or the $\rm J_s$-band flux were not used to constrain
the solutions. In this case, plausible solutions were characterized by
$\chi_{\nu}^2 < 5.41$, that corresponds to a probability of at least 0.001
for the only one degree of freedom for both starburst and Sbc-like models.
The new plausible solutions for starburst models allowed a slightly larger
parameter space but without major changes with the exception that a limited
number of plausible solutions with a $\chi_{\nu}^2$ $<$ 1 did exist now
(see Table \ref{tab:fits}).
Also for normal star-forming bulge$+$disk models the parameter space allowed
by the new plausible solutions became slightly larger
(see Table \ref{tab:fits2}); in particular, the bulge-to-total mass ratio
was unconstrained.
These new solutions spanned the whole range in central opacity,
the least opaque models ($\tau_V^{\rm c, 0} = 0.50$) having older ages
(1.5 -- 2.9 Gyr) than the most opaque ones (with $\tau_V^{\rm c, 0} = 16$
and an age of 1.0 -- 1.7 Gyr). Models with larger bulge-to-total mass ratios
tended to be younger, independent of the central opacity; however,
the stellar mass was still a few to several times $\rm 10^{9}~M_{\sun}$
overall. This time plausible solutions with a $\chi_{\nu}^2 < 1$ did exist
also for Sbc-like models, without major changes in terms of properties
of the stellar populations and mass of the system.

\begin{table*}
\caption{Results of Starburst Model Fits w/o $\rm J_s$-band data}
\label{tab:fits}
\centering
\begin{tabular}{ccccc}
\hline\hline
$\tau_V$ & age & Z & M$_{\star}$ & $\chi_{\nu}^2$ \\
& Gyr & $10^{-3}$ & $\rm 10^{9}~M_{\sun}$ & \\
\hline
0.25 -- 1.5 & 0.4 -- 2.0 & 0.3 -- 1.6 & 3.1 -- 4.9 & $< 5.41$ \\
1 & 0.8 -- 0.9 & 0.6 -- 0.7 & 3.5 -- 3.8 & $< 1.00$ \\
\hline
\end{tabular}
\end{table*}

\begin{table*}
\caption{Results of Sbc-like Model Fits w/o $\rm J_s$-band data}
\label{tab:fits2}
\centering
\begin{tabular}{cccccc}
\hline\hline
B/T$^{\mathrm{1}}$ & $\tau_V^{\rm c, 0}$ & age & Z & M$_{\star}$ & $\chi_{\nu}^2$ \\
& & Gyr & $10^{-3}$ & $\rm 10^{9}~M_{\sun}$ & \\
\hline
0.05 & 0.50 -- 16 & 1.0 -- 2.9 & 3.0 -- 8.5 & 3.6 -- 6.4 & $< 5.41$ \\
0.05 & 4, 16 & 1.3 -- 1.5 & 3.9 -- 4.6 & 4.8 -- 5.0 & $< 1.00$ \\
0.10 & 0.50 -- 16 & 1.0 -- 2.6 & 3.0 -- 7.7 & 3.3 -- 6.4 & $< 5.41$ \\
0.10 & 0.50 -- 8 & 1.5 -- 2.1 & 4.6 -- 6.3 & 4.4 -- 5.1 & $< 1.00$ \\
0.15 & 0.50 -- 16 & 1.0 -- 2.5 & 3.0 -- 7.5 & 3.9 -- 6.9 & $< 5.41$ \\
0.15 & 0.50, 4, 8 & 1.3 -- 1.8 & 3.9 -- 5.5 & 4.4 -- 4.9 & $< 1.00$ \\
0.20 & 0.50 -- 16 & 1.0 -- 2.3 & 3.0 -- 6.9 & 3.6 -- 6.8 & $< 5.41$ \\
0.20 & 0.50, 1, 4, 16 & 1.2 -- 1.9 & 3.6 -- 5.8 & 4.6 -- 5.6 & $< 1.00$ \\
\hline
\end{tabular}
\begin{list}{}{}
\item[$^{\mathrm{1}}$] Bulge-to-total mass ratio.
\end{list}
\end{table*}

\section{Star Formation Rate}

The previous plausible solutions give values of the SFR
equal to 3.1 -- $\rm 9.4~M_{\sun}~yr^{-1}$ (starburst models)
or 2.4 -- $\rm 4.1~M_{\sun}~yr^{-1}$ (normal star-forming, Sbc-like models),
the value of SFR decreasing as the time elapsed since the start
of star formation increases
\footnote{For a {\em different region
of the host galaxy GRB\,011121 containing the OT}, Greiner et al. (\cite{g03})
estimated values of the SFR from [OII] and $\rm H\alpha$ emission-line
diagnostics {\em at times when
the afterglow was present}. These values are: $\rm 1.2~M_{\sun}~yr^{-1}$
(SFR$_{OII}$) and 0.61 -- $\rm 0.72~M_{\sun}~yr^{-1}$ (SFR$_{H\alpha}$).
It is clear that these values do not refer to the whole galaxy
and are not corrected for the intrinsic extinction.}.
For the same models, the SFR per unit stellar mass is equal to
0.6 -- $\rm 2.9 \times 10^{-9}~yr^{-1}$ or
0.4 --$\rm 0.7 \times 10^{-9}~yr^{-1}$, respectively. Consistently,
for this subluminous galaxy ($\rm L_B/L^{\star}_B = 0.26$),
the SFR per unit luminosity is equal to
11.9 -- $\rm 36.1~M_{\sun}~yr^{-1}~(L_B/L^{\star}_B)^{-1}$
or 9.2 -- $\rm 15.8~M_{\sun}~yr^{-1}~(L_B/L^{\star}_B)^{-1}$.

These values of the SFR per unit stellar mass are high compared to those
of simulated galaxies in Courty et al. (\cite{cour04}), in agreement
with their conclusion that the GRB-host galaxies
are identified as the most efficient star-forming objects. Other GRB-host
galaxies have high values of the SFR per unit luminosity (cf. Christensen
et al. \cite{chri04a}), though not as high as our estimates. Recent calculations by Gorosabel et al. (\cite{gor05}) and Sollerman et al.
(\cite{sol05}) give similar values of the extinction-corrected SFR
per unit luminosity for the host galaxies of the two low-redshift
GRB\,030329 and GRB\,031203.

Finally, we compared the values obtained for the SFR per unit
galaxy stellar mass of the host galaxy of GRB\,011121 with those
of observed galaxies selected from the MUNICS and FORS deep field surveys
(Bauer et al. \cite{bau05}) in the same redshift range $0.25 < z < 0.4$
as the previous GRBs and GRB\,011121 itself. The values of the specific SFR
(SSFR) given by Bauer et al. (\cite{bau05}) were determined from the [OII]
line flux without any correction for dust extinction. This comparison confirms
that the host galaxy of GRB\,011121 is among the galaxies with highest
specific SFR at these redshifts even after allowing for an extreme correction
factor of 10 for the SSFRs given by Bauer et al. (\cite{bau05}).

\section{Summary}

The existence of high-resolution imaging in 5 broad-band, optical
and near-infrared filters with HST and VLT/ISAAC for the host galaxy
of GRB\,011121 (at $z = 0.36$) allows a detailed study of both the morphology
and the spectral energy distribution of this galaxy. Multi-band,
high signal-to-noise ratio, high-resolution imaging of GRB host galaxies
is still a luxury, only affordable for the brightest and most nearby galaxies.

Firstly, we find that the surface brightness profile of the host galaxy
of GRB\,011121 is best fitted by a Sersic law with index
$n \sim 2$ -- 2.5 and a rather large effective radius ($\sim$ 7.5 kpc).
Together with the F450W - F702W colour image, this suggests that this galaxy
is either a disk-system with a rather small bulge (like an Sbc galaxy),
or one hosting a central, dust-enshrouded starburst.

At variance with previous studies on GRB host galaxies, we combine
stellar population models and Monte Carlo calculations of radiative transfer
to reproduce the observed SED. Furthermore, we make use of the morphological
information to constrain these models. Plausible solutions meeting all
the morphological and/or photometric constraints indicate that the host galaxy
of GRB\,011121 has a stellar mass of a few to several times
$\rm 10^9~M_{\sun}$, stellar populations with a maximum age ranging
from 0.4 to 2 Gyr, and a bolometric luminosity-weighted metallicity in stars
(of the disk, in case) ranging from 1 to 29 per cent of the solar value.

In particular, normal star-forming, Sbc-like models provide plausible solutions
pointing to a system as massive as 4.9 -- $\rm 6.9 \times 10^{9}~M_{\sun}$,
with a bulge-to-total mass ratio equal to 0.15, an age of 1.3 -- 1.9 Gyr,
and a bolometric luminosity-weighted metallicity in stars of the disk
equal to 20 -- 29 per cent solar. On the other hand, starburst models provide plausible solutions biased towards a lower stellar mass
(3.1 -- $\rm 4.8 \times 10^{9}~M_{\sun}$), a younger age (0.4 -- 2.0 Gyr)
and a much lower metallicity (1 -- 8 per cent solar). As for the opacity,
normal star-forming, Sbc-like models indicate the host galaxy of GRB\,011121
as a system with a central opacity $\tau_V^{\rm c, 0}$ in the range 2 -- 16,
i.e. larger than the central opacity of local disks (0.5 -- 2, see Kuchinski
et al. \cite{kuc98}). Nevertheless, the attenuation along the line of sight
is moderate ($\rm A_V = 0.12$ -- 0.57 mag) on the scale of the system
since the host galaxy of GRB\,011121 has a low inclination (18 degrees).
On the other hand, starburst models suggest this galaxy to be nearly as opaque
($\tau_V = 0.5$ -- 1.5) as local starburst galaxies (with $\tau_V \sim 1.5$,
see Gordon et al. \cite{gcw97}), the attenuation along the line of sight being
$\rm A_V = 0.27$ -- 0.76 mag on the scale of the system.

The SFR per unit stellar mass is equal to
0.6 -- $\rm 2.9 \times 10^{-9}~yr^{-1}$ (starburst) or
0.4 --$\rm 0.7 \times 10^{-9}~yr^{-1}$ (normal star-forming galaxy),
while the SFR per unit luminosity is equal to
11.9 -- $\rm 36.1~M_{\sun}~yr^{-1}~(L_B/L^{\star}_B)^{-1}$
or 9.2 -- $\rm 15.8~M_{\sun}~yr^{-1}~(L_B/L^{\star}_B)^{-1}$, respectively.

This large (effective radius of $\sim$ 7.5 kpc) but subluminous
($\rm L_B/L^{\star}_B = 0.26$) galaxy exhibits a specific SFR that is larger
than that of the average galaxy at the same redshift (e.g. Bauer et al.
\cite{bau05}) but consistent with the values determined for two other blue, low-metallicity, low-$z$ GRB host galaxies (i.e. GRB\,030329
and GRB\,031203, see Gorosabel et al. \cite{gor05}, Sollerman et al.
\cite{sol05}). Therefore, we conclude that the host galaxies of GRB\,011121
and, possibly, GRB\,030329 and GRB\,031203 are cought at relatively
early phases of their star formation histories.

\begin{acknowledgements}
We thank to the anonymous referee for extensive comments that helped to improve the paper. AKY acknowledges support from the International Max-Planck Research School 
(IMPRS) on Astrophysics. MS acknowledges Sonia Temporin for a lively 
discussion. EP is grateful to the MPE for hospitality and support. AR 
acknowledges support and collaboration within the EU RTN Contract
HPRN-CT-2002-00294.
\end{acknowledgements}

\end{document}